# Interfacial charge-transfer in 3*d*/5*d* complex oxide heterostructures

[1,#]Arun Kumar Jaiswal, [2]Di Wang, [1]Ji Soo Lim, [1]Shruti Roy, [3]Fabrice Wilhelm, [2]Vanessa Wollersen, [3]Andrei Rogalev,[1]Matthieu Le Tacon, and [1]Dirk Fuchs[*]

[1]Karlsruhe Institute of Technology, Institute for Quantum Materials and Technologies, Kaiserstr. 12, 76131 Karlsruhe, Germany;
[2]Karlsruhe Institute of Technology, Institute of Nanotechnology and Karlsruhe Nano Micro Facility (KNMFi), Kaiserstr. 12, 76131 Karlsruhe, Germany;
[3]European Synchrotron Radiation Facility (ESRF), F-38043 Grenoble, France;

#: *current address*:

Department of Quantum Matter Physics, University of Geneva, CH – 1211 Geneva 4, Switzerland

[*]: *contact*:

dirk.fuchs@kit.edu

Interfacial charge transfer (ICT) provides a powerful route to engineer electronic phases in correlated oxide heterostructures, yet predictive design principles remain elusive. Here, we systematically investigate superlattices composed of the 5*d* spin–orbit coupled semimetal $SrIrO_3$ and a series of correlated 3*d* perovskites ($LaMnO_3$, $LaFeO_3$, and $LaCoO_3$), thereby establishing a quantitative framework for ICT across 3*d*/5*d* interfaces. Combining Ir $L_{2,3}$ edge X-ray absorption spectroscopy with spatially resolved electron energy loss spectroscopy, we estimate the electron depletion in the $SrIrO_3$ layers and the corresponding electron accumulation in the 3*d* layers. The magnitude of ICT follows the trend expected from electronegativity mismatch, supporting electronegativity-driven band alignment as the dominant contribution in this series. Beyond interfacial doping, we find that strong 3*d*–5*d* hybridization induces a low-spin to high-spin conversion in the cobaltate layers, hinting to interface-controlled spin-state engineering without chemical substitution. These results establish electronegativity mismatch as a predictive design parameter for correlated oxide interfaces and provide a materials platform for tailoring band filling, orbital hierarchy, and spin configurations in quantum oxide heterostructures, paving the way towards advanced oxide electronics and next-generation information technologies.

## 1. Introduction

Complex oxides present one of the most important class of functional materials for electronic applications.[1] The strong electron correlation in the 3*d* transition metal oxides (TMOs) can result, for example, in magnetism, ferroelectricity, superconductivity or even multiferroic ordering phenomena,[2] whereas the 4*d* and 5*d* TMOs are well known for their strong spin-orbit coupling and topological properties.[3] The connected network of metal-oxygen bonds provides perfect conditions to manipulate the functionalities of TMOs by *e. g.*, pressure or strain,[4–10] which has made epitaxial TMO thin films highly attractive in solid-state physics over the past decades. Meanwhile, artificial heterostructures or superlattices (SLs) consisting of TMOs with different functionalities can be grown with atomic precision which provides a versatile playground for the design of novel materials.[11,12] The emergence of new phenomena at the interface of TMO heterostructures has stimulated scientific interest, as the reconstruction of the charge, spin and orbital degrees of freedom can lead to exotic quantum states of fundamental and practical relevance.[13–15] Here, the combination of 3*d* and 5*d* TMOs in the form of heterostructures have gained special interest since strong electron correlation and spin-orbit coupling coexist at the heterointerface. For example, proximity-induced magnetism has been observed in the weakly electron-correlated iridates at the interface with manganates[16,17] or cobaltates.[18] On the other side, transfer of spin-orbit coupling from the 5*d* iridates to the 4*d* TMO $SrRuO_3$ has been discussed alike.[19,20]

Charge transfer is often observed across the interface of TMO heterostructures [21–30] making it as a key ingredient for the observation of emergent phenomena such as proximity induced magnetism[17,18,31] or interfacial conductivity.[14,32,33] Hence, understanding ICT mechanism is essential to improve controllability of key functionalities of heterostructures. Usually, ICT in TMO heterostructures is limited to a few unit cells from the interface.[34–37] Therefore, short-period SLs are highly preferred to study ICT. The possibility to tune intrinsic doping or covalence across the interface provides promising strategies to engineer new quasi two-dimensional states in correlated oxides.

The continuity of electronic states in the oxygen matrix of complex corner-sharing TMOs forces band-alignment at the heterointerface, leading to a potential gradient that triggers electron- or hole- transfer across the interface.[38] In contrast to semiconductors, electron affinity or work function of TMOs depend on the surface termination and therefore are not suitable to describe band alignment at such interfaces.[39] In perovskite ($ABO_3$) heterostructures, a chemical potential mismatch is often originated from differences in (i) polarity, as in $LaAlO_3$/$SrTiO_3$;[32] (ii) the valence state of the *B*-site ion, as in $LaMnO_3$/$SrMnO_3$;[40] or (iii) the electronegativity, which increases across the 3*d* TMOs with increasing *d*-shell filling.[25] If these differences are not compensated by *e. g*., structural reconstruction, defects or adsorbates, charge transfer will occur across the interface.[41] Recently, a simple framework has been proposed to predict band-alignment and charge transfer in complex TMOs.[38] Based on the continuity of MO-related valence states across the interface, the sign and magnitude of the intrinsic charge transfer $\Delta n_e$ are determined by the energy difference between the local oxygen 2*p* states, $\Delta\varepsilon_p$. Generally, $\varepsilon_p$ increases across the 3*d* TMOs with increasing *d*-electron filling or oxidation state within the same period or decreasing group number of the periodic table of elements.[38] In heterostructures, the component with the lower $\varepsilon_p$ donates electrons to the one with higher $\varepsilon_p$. Accordingly, within 3*d*/5*d* TMO SLs electrons are expected to be transferred from the Ir side to the 3*d* TM.

However, compressive epitaxial strain and M-O hybridization can substantially modify the local $\varepsilon_p$ and consequently influence ICT.[38] In particular, the role of M-O covalency has been demonstrated in rare-earth nickelate $ANiO_3$ heterostructures, where variations in Ni–O covalency increases with *A*-site ionic size affecting electron-affinity in $ANiO_3$/$GdTiO_3$ heterostructures.[36,42]

Even though meanwhile reported in several 3*d* TMO heterostructures, a systematic quantitative analysis of the charge transfer and its driving mechanism is still missing. Here, we present a systematic study of the charge transfer across the interface between a strongly electron correlated 3*d* TMO ($LaMnO_3$ (M), $LaFeO_3$ (F), or $LaCoO_3$ (C)) and the spin-orbit coupled semi-metallic 5*d* TMO $SrIrO_3$ (I). To this end, short period SLs of the form $[I_4X_4]_5$ with X = M, F, and C, a layer thickness of 4 monolayers (MLs) and a periodic repetition of 5 were produced by pulsed laser deposition. Valence state of the *B*-site ions are investigated to study ICT by using X-ray absorption spectroscopy (XAS) and spatially resolved electron energy loss spectroscopy (EELS). The magnitude of the ICT scales linearly with the difference in electronegativity between the TMO layers and identifies electronegativity-mismatch as the driving mechanism for ICT. Additionally, 3*d*–5*d* hybridization induces a low-spin (LS) to high-spin (HS) conversion in the cobaltate layers, hinting to interface-controlled spin-state engineering without chemical substitution. Key parameters governing ICT in 3*d*-5*d* heterostructures are established providing a material design strategy for tailoring quasi two-dimensional states and emergent interfacial functionalities in complex oxides.

## 2. Results and Discussion

### 2.1. Structural properties of 3*d*/5*d* oxide superlattices

Epitaxial thin film superlattices (SLs) of the form $[I_4X_4]_5$ were prepared by pulsed laser deposition on (001) $SrTiO_3$ (STO) substrates and subsequently characterized by X-ray diffraction (XRD) and high-resolution scanning transmission electron microscopy (HR-STEM), see Fig. 1. For symmetric diffraction conditions, 2θ/ω-scans indicate cube-on-cube growth of the perovskite layers with pseudo-cubic $(001)_{pc}$ growth orientation. First-order superlattice reflections appear to the left and the right of the main Bragg-peak ($2\times\sin(\theta)/\lambda = 2/(c_I+c_X) \pm 1/\Lambda$) and reveal periodicity of $\Lambda = 4\times(c_I+c_X)$, where $c_I$ and $c_X$ correspond to the pseudo-cubic out-of-plane lattice parameters of I and X, respectively, and $\lambda$ = 1.5418 Å to the wavelength of Cu *K*α radiation.[43] $c_{I,X}$ were extracted from single-layer ($I_{10}$ and $X_{10}$) and bi-layer ($I_{10}X_{10}$) samples, which display the same in-plane lattice parameter and strain state as the $[I_4X_4]_5$ SLs (see SI). The pseudo-cubic in-plane lattice parameters $a_{I,X}$ were deduced from reciprocal space maps on asymmetric $(204)_{pc}$-type reflections (see SI) and found to be identical to that of the used STO substrate ($a_I = a_X$ = 3.905 Å) showing coherent and homogeneously strained epitaxial growth of the perovskite layers. In- and out-of-plane film lattice parameters and corresponding epitaxial strain, $\varepsilon_{xx} = (a-a_{pc})/a_{pc}$ and $\varepsilon_{zz} = (c-a_{pc})/a_{pc}$, of the layers are summarized in Table 1. Pseudo-cubic bulk lattice parameters have been deduced from the orthorhombic lattice parameters[44,45] by $a_{pc} = 1/3\times(a_{or}/\sqrt{2}+b_{or}/\sqrt{2}+c_{or}/2)$. Layers of I, M, and F are under compressive strain ($\varepsilon_{xx} < 0$), whereas C experiences tensile strain. For isotropic materials, a linear strain relation of TMOs usually results in a Poisson ratio $\nu = -\varepsilon_{xx}/\varepsilon_{zz} \approx 0.3$,[46] which is only obtained for the moderately strained F-layers ($\nu$ = 0.29). Distinct deviations are observed for larger in-plane lattice strain, $\varepsilon_{xx}$, hinting to an anisotropic Young's modulus or lattice relaxation along the out-of-plane growth-direction.[46,47] The octahedral rotation pattern of the different layers could not be verified within experimental resolution but likely differs from the bulk as well, as observed in $[I_{10}C_{10}]$ bi-layers.[18,48]

The interfacial structural properties of the SLs were analyzed by cross-sectional high-angle annular dark-field STEM (HAADF-STEM) images, shown in Fig. 1(c-e). The HAADF-STEM images indicate atomically sharp I-X interfaces with atomic interdiffusion not more than 1 ML. Atomically resolved elemental maps were acquired by spectrum imaging combining STEM and energy dispersive X-ray spectroscopy (STEM-EDS spectrum imaging). Transition metal oxide layers close to the interface appear to be sharper compared to the alkaline/rare earth metal oxide layers. More specifically, quantitative STEM-EDS mapping of the SL cross-section (see SI, Fig. S3) reflect the element proportion and indicates dominant SrO-termination of the I-layer with some admixture of LaO – possibly due to incomplete SrO-layers, whereas the X-layer for X = M, F, and C shows nearly perfect LaO-termination.

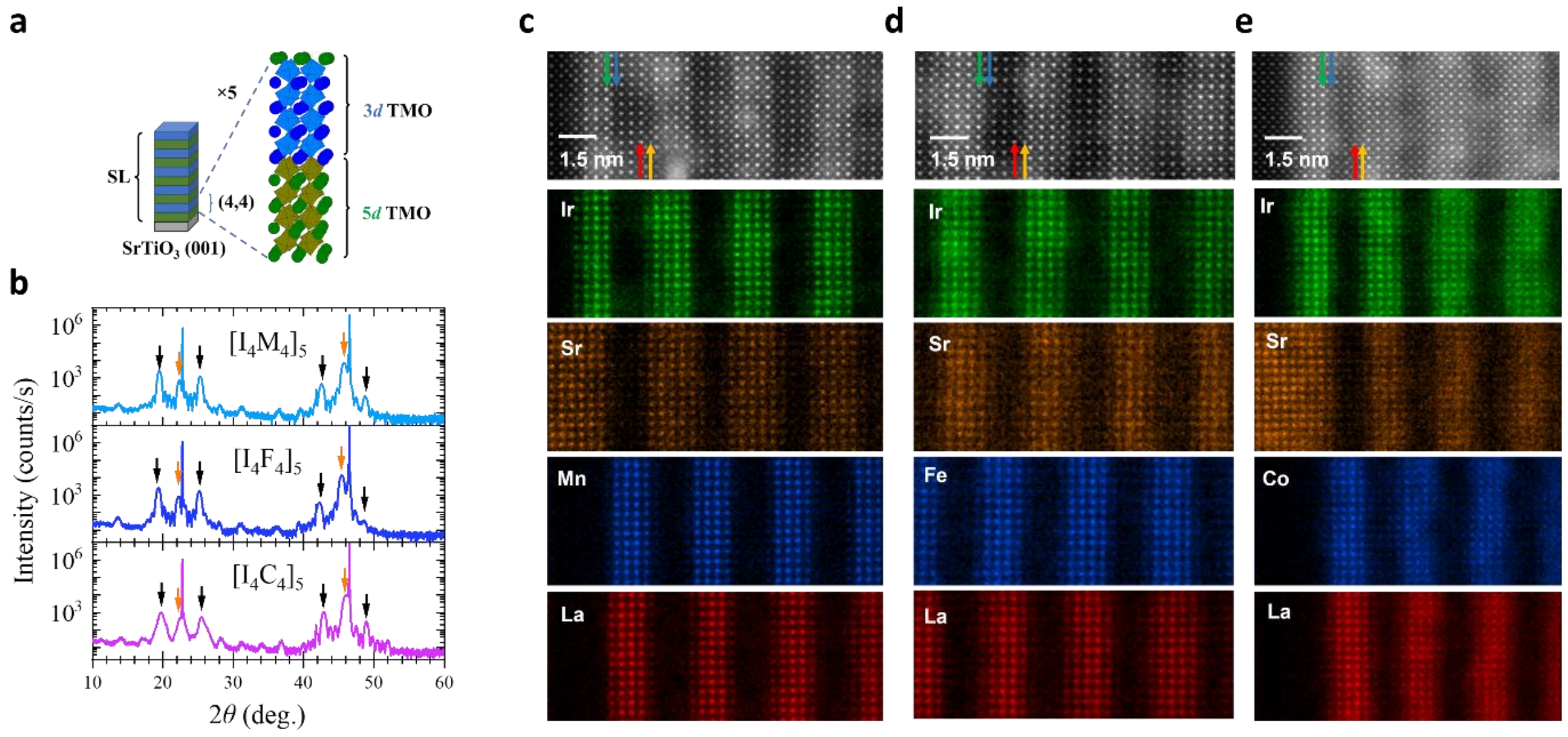

**Figure 1**. (a) Sketch of the 3*d*/5*d* SL $[I_4X_4]_5$. The modulation length comprises 4 monolayers (MLs) of the 3*d* (X) and 4 MLs of the 5*d* (I) perovskite with a total repetition of 5. (b) Symmetric 2θ/ω X-ray diffraction scans of the $[I_4X_4]_5$ (X = M, F, and C) SLs. Main SL Bragg peak ($SL_0$) and first-order SL peaks, $SL_{-1}$ (left) and $SL_{+1}$ (right) are indicated by orange and black arrows, respectively. (c-e) Transmission electron microscopy on $[I_4X_4]_5$: HAADF-STEM image (top) and element specific maps from STEM-EELS spectrum imaging (below). Atomic layers are indicated by color and colored arrows: $IrO_2$ (green), SrO (orange), $MnO_2$/$CoO_2$/$FeO_2$ (blue), and LaO (red). The images display SrO (LaO)-termination of the I (M, F, and C) -layer.

**Table1**. Film lattice parameters, bulk pseudo-cubic lattice parameter, lattice strain and tetragonal distortion of the SL perovskite layers grown on (001) $SrTiO_3$.

| Layer | *a* (Å) | *c* (Å) | $a_{pc}$ (Å) | $\varepsilon_{xx}$ (%) | $\varepsilon_{zz}$ (%) | *c*/*a* |
|---|---|---|---|---|---|---|
| $SrIrO_3$ | 3.905 ±0.005 | 4.07 ±0.006 | 3.943 | -0.96 | 3.22 | 1.04 |
| $LaMnO_3$ | 3.905 ±0.005 | 4.03 ±0.014 | 3.938 | -0.84 | 2.33 | 1.03 |
| $LaFeO_3$ | 3.905 ±0.005 | 4.01 ±0.001 | 3.922 | -0.43 | 2.24 | 1.03 |
| $LaCoO_3$ | 3.905 ±0.005 | 3.77 ±0.002 | 3.820 | +2.23 | -1.31 | 0.96 |

## 2.2. X-ray absorption spectroscopy at the Ir *L*-edges of the 5*d* TMO

In the $[I_4X_4]_5$ SLs the ICT was studied by valence state analysis of the 3*d* and 5*d* TM. To this end XAS and EELS measurements at the TM *L*-edges were carried out. Particularly for 3*d*/5*d* TM (2*p* → 3*d*/5*d* transitions), *L*-edges XAS reveals valence states through energy shifts of the absorption edge and changes in the "white line" (WL) intensity and spectral shape, reflecting the 3*d*/5*d* empty density of states, allowing precise differentiation of oxidation states and even multiple coexisting states in a single spectrum.[49] Quantitative information on the charge carrier depletion in the I-layer was obtained by Ir $L_{2,3}$-edges XAS. The Ir $L_2$-edge is sensitive to transitions involving $5d_{3/2}$ (*i.e.* $J_{eff}$ =3/2) holes, while the $L_3$-edge is related to both $5d_{5/2}$ ($J_{eff}$ =1/2 and the crystal field $e_g$ manifolds) and $5d_{3/2}$ final states. The sum of the integrated intensities of the $L_2$ and $L_3$ WLs is proportional to the number of unoccupied states and hence the number of holes $n_h$ in the 5*d* shell of Ir. Therefore, WL analysis provides information on the Ir valence state and charge transfer. For a quantitative WL peak-area analysis, the XAS spectra were normalized after careful background subtraction, see SI.

In Fig. 2a we show the normalized Ir $L_{2,3}$-edges XAS spectra for the SLs in comparison to a SIO reference sample $I_{20}$, *i. e.*, 20 MLs of SIO on STO. The mean number of holes $n_h$ on the Ir-site which we deduced from the spectra is shown in Fig. 2b. For $I_{20}$, an Ir $4^+$ ($5d^5$) valence state ($n_h = 5$) was assumed. For all SLs, $n_h$ is larger than 5 showing an increase of the 5*d*-holes and hence electron transfer from the Ir-site of the I-layer to the X-layer. With respect to $I_{20}$, $n_h$ increases for $[I_4M_4]_5$, and $[I_4F_4]_5$ and is largest (5.35) for $[I_4C_4]_5$, yielding $\Delta n_h = 0.35$. $\Delta n_h$ seems to increase with increasing the occupation of the 3*d* shell of the X-layer. With respect to the reference sample $I_{20}$, the SLs display relative increase of the number of the 5*d* holes per Ir-site of about 3.2% for $[I_4F_4]_5$ which roughly doubles to 6.9% for $[I_4C_4]_5$. The observed electron depletion by a factor of about 2.2 is well consistent with Hall measurements, see below and SI. Because of the semi-metallic behavior and low resistivity of the I-layer, electron depletion is assumed to be homogeneous throughout the I-layers.

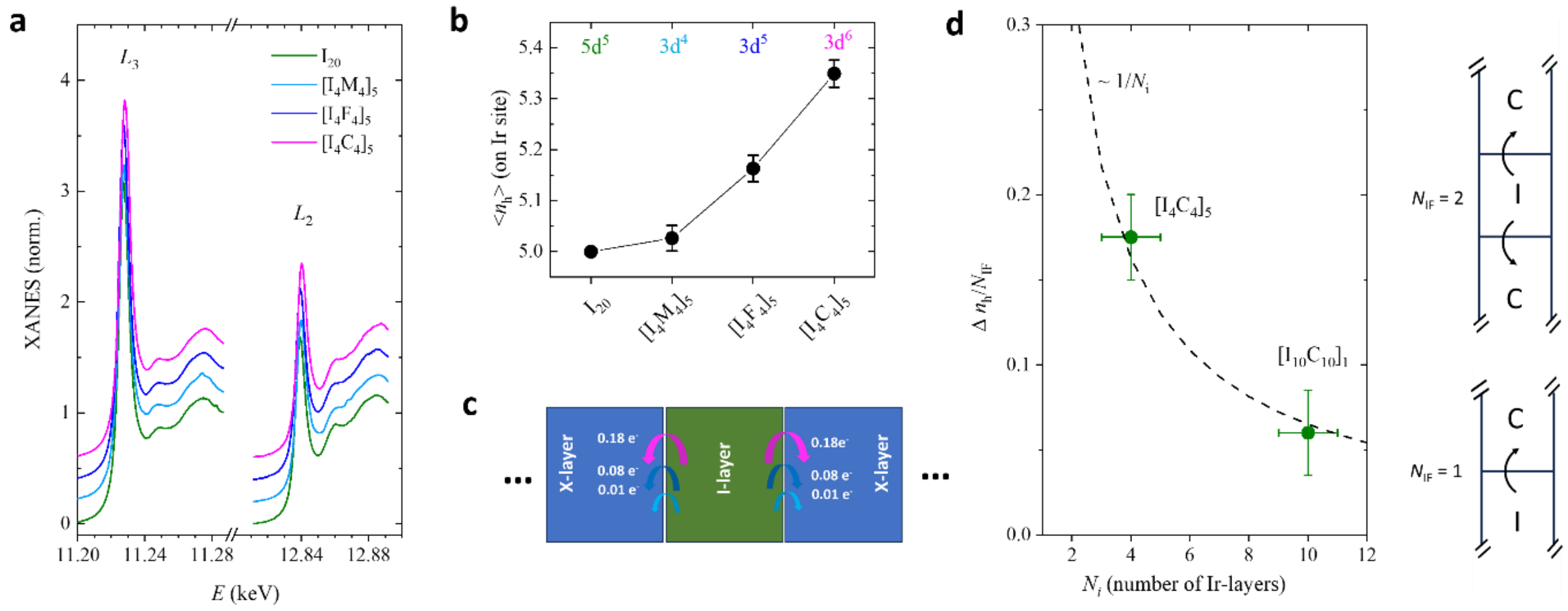

**Figure 2**. (a) Normalized Ir $L_{2,3}$ XANES of SIO ($I_{20}$) and the $[I_4X_4]_5$ SLs. Measurements were taken at grazing beam incidence (15°) at $T = 20$ K. Spectra of SLs are shifted vertically for clarity. (b) The mean number of holes $n_h$ on the Ir-site for $I_{20}$ (SIO) and the $[I_4X_4]_5$ SLs extracted from the sum area of the Ir $L_2$ and $L_3$ white lines (WLs). The sum of the WL integrated intensities of $I_{20}$ is fixed to $n_h = 5$, *i. e.*, an Ir $^{4+}$ ($5d^5$) valence state. (c) Electron transfer per Ir from the I- to the X-layer as deduced from (b). Color code is same as before. (d) The mean hole number on the Ir site per interface $\Delta n_h/N_{IF}$ which scales with the inverse of the number of Ir MLs ($N_i$) in the I-layer (see dashed line). The situation for the SL ($N_{IF} = 2$) and the bi-layer ($N_{IF} = 1$) is sketched on the right.

Evidently, ICT occurs when the I-layer is in direct contact to the X-layer so that it is expected to happen at both interfaces of the I-layers equally. Assuming that total hole and electron accumulation are same ($\Delta n_h = \Delta n_e$), electron transfer to the right and left interface of the I-layer then amounts to $\Delta n_h/2$, as illustrated in Fig. 2c. The $[I_4C_4]_5$ SL ($N_{IF} = 2$) exhibits a hole accumulation of $\approx 0.18$ per interface. In contrast, a thicker $[I_{10}C_{10}]_1$ bi-layer ($N_i = 10$, $N_{IF} = 1$) shows $\Delta n_h = 0.06$ (see Fig. 2d). The hole-accumulation per interface $\Delta n_h/N_{IF}$ scales perfect with $1/N_i$ so that $\Delta n_h \times N_i/N_{IF}$ is nearly constant for different types of heterostructures with certain X, which demonstrates the interfacial character of the charge transfer. The extracted charge transfer fits very well to values reported from density functional theory (DFT) calculations.[38]

Electron depletion in the semi-metallic I-layer also becomes visible in electronic transport because of the short modulation length or layer-thickness of the SLs. However, quantification of ICT from electronic transport is challenging. A layer thickness of only a few MLs can also lead to electron confinement and may mask ICT in resistance measurements by a decrease of charge carrier mobility (see SI). Analysis of the charge carrier depletion or accumulation of each layer from the ordinary Hall effect may become even impossible, when both constituent layers contribute to the conductivity. Contributions to the total conductivity cannot be fully excluded for SLs with X = M, in contrast to F and C, where conductivity is only related to the I-layer, see SI. In comparison to a single I-layer sample ($I_{10}$) the nearly linear $B$-dependence of the Hall resistance $R_{xy}$ indicate electron depletion by about 8% for $I_{10}F_{10}$ heterostructures which increases by a factor of 2.4 to 19% for and $I_{10}C_{10}$. Assuming a one-type carrier model, these changes of the Hall slope indicate electron reduction in the $SrIrO_3$ layer very similar as deduced from Ir $L_{2,3}$ XAS. However, because $SrIrO_3$ exhibits multiband and nearly compensated electron-hole transport, and because mobility changes cannot be excluded, we use these data only as qualitative supporting evidence. The quantitative determination of charge transfer relies on the Ir $L_{2,3}$ XAS and EELS analyses.

## 2.3. Spatially resolved electron energy loss spectroscopy at the 3*d* TM

In contrast to Ir, the $L_{2,3}$ absorption edges of the 3*d* TMs occur at much lower energies allowing an analysis of the valence state by spatially resolved electron energy loss spectroscopy (EELS) measurements in TEM, see SI. This has become a crucial tool for quantifying the local valence states of many 3*d*-TMs, including Fe, Mn, and Co.[50–52] To this end, STEM-EELS spectrum imaging were taken on the SLs for 3*d* TM ions (Mn, Fe, and Co), see SI. The so obtained 3*d* TM $L_{2,3}$ EELS spectra were integrated, and the $L_3/L_2$-intensity ratio (branching ratio BR) was extracted for the individual MLs of the SLs, see Fig. 3. More details on data evaluation are given in SI. EELS and XAS are complementary methods for analysis of the BR with EELS providing higher spatial- and XAS higher spectral resolution. Both offer similar electronic structure information and effectively probe unoccupied states using the spin-orbit sum rule.

The $L$-edge spectra of 3*d* and 5*d* TMs differ substantially due to their distinct orbital energies and spin-orbit coupling. In 3*d* TMs, stronger electron-electron interactions give rise to pronounced multiplet effects and deviations from the statistical $L_3/L_2$-intensity ratio of 2:1, (four $2p^{3/2}$ and two $2p^{1/2}$ electrons). Similar to XAS, deviations of BR in EELS reflect the occupation of 3*d* orbitals. From the theoretical model calculations, the HS state displays larger BR than the LS state, which makes the BR also sensitive to the spin state of the 3*d* TM ion.[53]

In comparison to the 3*d* TMs, the 5*d* Ir $L_{2,3}$ edges show sharper peaks due to weak electron-electron interaction, allowing a single-particle picture and facilitating quantitative valence state analysis. For 3*d* TMs, pronounced multiplet effects require comparison of the BR with standard compounds with well-known valence- and spin states. Spatially resolved EELS spectra are shown for the $[I_4X_4]_5$ SLs in Fig. 3b-d for X = M, F, and C. Interestingly, the BR for X = M, F and C is comparable for all four MLs within experimental resolution, indicating a homogeneous valence state throughout the X-layer, as expected from the spatially limited ICT in TMO heterostructures.[34–37]

The robust HS state of the TM in $LaMnO_3$ and $LaFeO_3$ allows to compare BRs obtained from $[I_4M_4]_5$ and $[I_4F_4]_5$ SLs with standard reference compounds and epitaxial single layer films of X with similar film thickness (20 ML), see SI. [54–57] Differences between single layer M and F and the corresponding SL are within experimental resolution, however clearly visible for X = C, which can be interpreted by changes due to ICT since in-plane lattice spacing and epitaxial strain of the single layer C film and SL are the same. The BR of around 4.0 for the $[I_4C_4]_5$ SL indicates valence state considerably lower than 3+, agreeing well with the value deduced from Ir $L_{3,2}$ XAS (<val> = +2.65). For X = M and F, the estimated mean valences are +2.97±0.05 and +2.83±0.15, respectively, corresponding to nominal electron accumulations of 0.03±0.05 and +0.17±0.15 electrons per ion. These values are consistent with the charge transfer of [54–57]0.02 and 0.16 electrons per ion inferred from Ir $L_{2,3}$-edge XAS. However, since the differences between the superlattices and the corresponding reference films are within the experimental resolution of EELS, the measurements do not independently resolve such small charge transfers.

The BR of the $[I_4C_4]_5$ superlattice is substantially enhanced compared with the epitaxially strained $LaCoO_3$ reference film, pointing to a pronounced modification of the Co electronic configuration. The BR depends on both the valence and spin state of Co, with reported reference values of approximately 4.8 for HS $Co^{2+}$ and 2.6 for LS $Co^{3+}$.[52] An enhanced HS population is also known to increase the BR of $Co^{3+}$.[58] Indeed, the BR of the strained $LaCoO_3$ reference film (3.59 ± 0.17) exceeds that of bulk $LaCoO_3$ (3.35), consistent with the increased HS population reported for epitaxially strained cobaltate films.[59–64]In the superlattice, however, the larger BR of approximately 4.0 may reflect both the reduction of the nominal Co valence due to ICT and a redistribution of the Co spin states. Since the limited energy resolution of our EELS measurements does not allow a detailed multiplet analysis, the relative contributions of these effects cannot be determined unambiguously. Nevertheless, the pronounced modification of the Co L-edge branching ratio suggests that interfacial electronic reconstruction extends beyond a simple rigid-band description.

In summary, electron depletion in the I-layer and electron accumulation in the X-layer are well comparable and consistently document ICT in the $[I_4X_4]_5$ SLs. The short period of the SL ($i$ = 4) ensures a spatially homogeneous accumulation zone in the insulating X-layer facilitating a quantitative comparison of the ICT with respect to the 3$d$ TM. Beyond charge transfer, the enhanced Co branching ratio may also reflect modifications of the spin-state distribution associated with interfacial hybridization, although the present measurements do not allow these contributions to be disentangled.

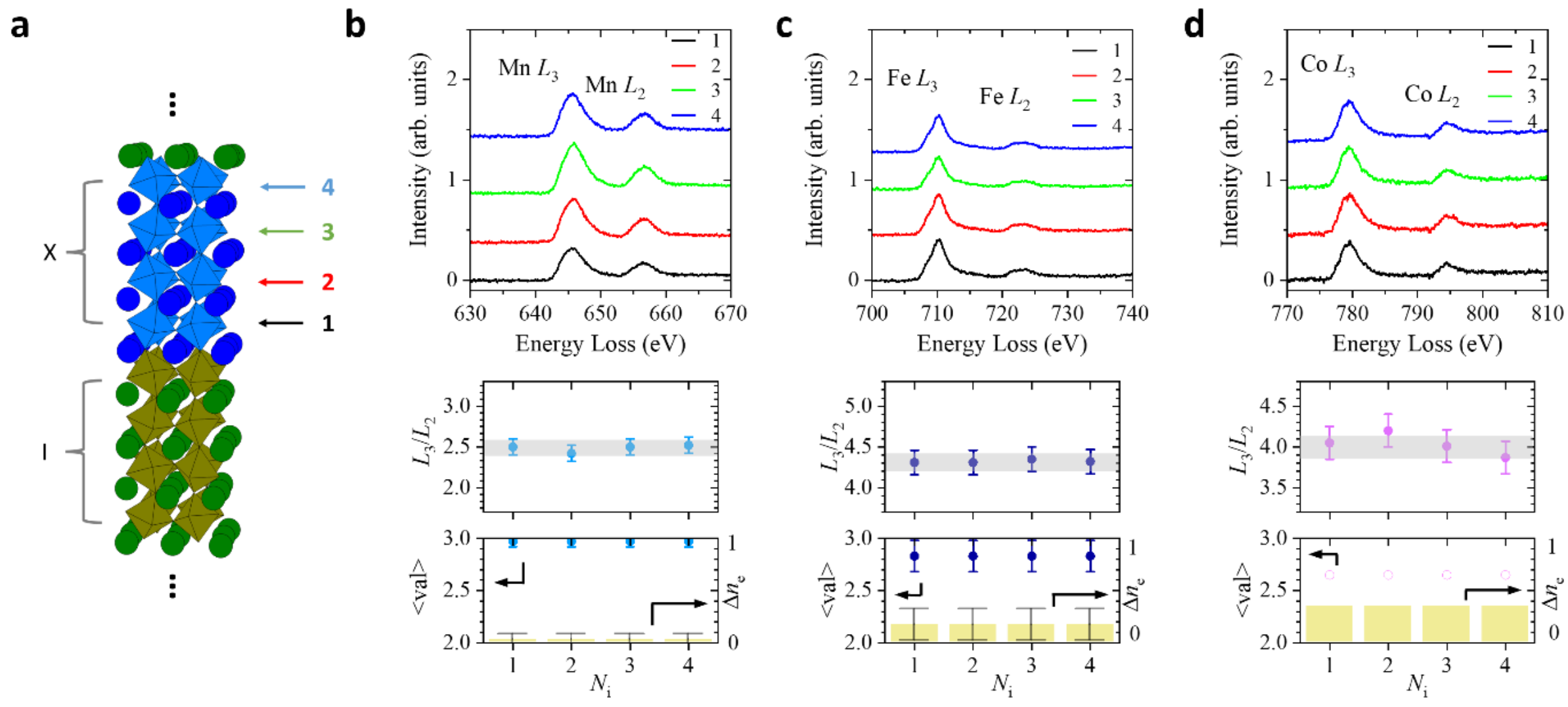

**Figure 3**. Spatially resolved EELS on $[I_4X_4]_5$ heterostructures. (a) Schematic of the $N_i$ ($i$ = 1-4) MLs of the X-layer analyzed by EELS measurements. (b-d) EELS spectra (top), $L_3/L_2$ branching ratio (middle) and mean valence number <val> of the $N_i^{th}$ ML as deduced from the spectra for X = M, F, and C (bottom, left scale). The mean electron accumulation per ion for each ML is indicated by yellow bars (right scale). Mean value of BR for the X-layer is indicated by grey bar.

## 2.4. Charge transfer mechanism at the interface

To identify the origin of ICT, we consider possible charge-transfer mechanisms at oxide perovskite interfaces and their relevance to $[I_iX_i]_m$ SLs. In Fig. 4(a-d), we have schematically illustrated the most prominent scenarios,[25] *i. e.*, ICT driven by differences in (i) the *B*-site valence state, (ii) the polarity, or (iii) the electronegativity of the neighboring layers.

For mechanism (i), in $ABO_3$ and $A'BO_3$ compounds with different *B*-site valence state, *e. g.*, 4+ and 3+, charge transfer may occur from $B^{3+}$ to $B^{4+}$ to smooth the occupancy discontinuity. In the present case, however, the *B*-site ions differ, and experimental results show the opposite direction of ICT, *i. e.*, from $Ir^{4+}$ to the 3*d* TM $M^{3+}$ indicating that this mechanism is not dominant.

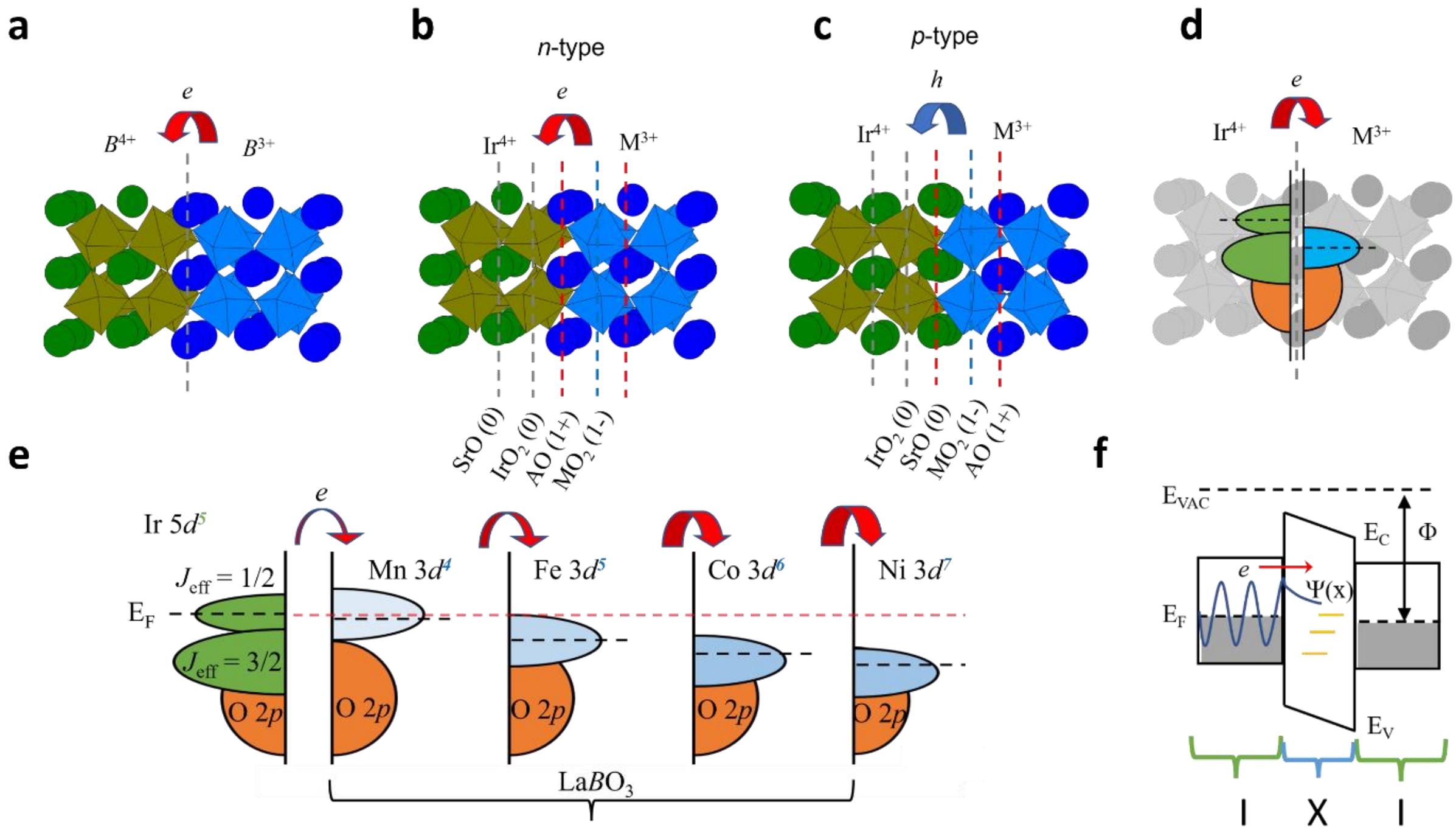

**Figure 4**. Most prominent types of charge transfer mechanisms at the interface of perovskite-like TMO heterostructure. (a) Occupancy difference (type-i): An abrupt change of the valence state of *B*-site ion across the interface can drive electron flow from $B^{3+}$ to $B^{4+}$ to smoothen occupancy discontinuity. (b) Polarity difference (type-ii): For 001-oriented $A^{2+}M^{4+}O_3^{6-}/A^{3+}M^{3+}O_3^{6-}$ perovskite heterostructures, the$[A^{2+}O^{2-}]/[M^{4+}O_2^{4-}]$ and $[A^{3+}O^{2-}]/[M^{3+}O_2^{4-}]$ layering result in a so-called polar-mismatch at the interface. To avoid the formation of a large built-in electric field, depending on the layer termination ($MO_2$ or AO) half an electron (n-type) or (c) hole (p-type) is transferred from right to the left. (d) Electronegativity difference (type-iii). Usually, electronegativity of transition metals increases from 5*d* to 3*d* within one group of the periodic table. In TMO heterostructures, oxygen states usually align across the interface, so that the electronegativity difference directly translates into a contribution of the chemical potential difference $\Delta E_F$ which is compensated by charge transfer to the TMO with the lower $E_F$., *i. e.*, the 3*d* TMO. (e) Charge transfer at the interface of Ir5*d*/X3*d* (X = Mn, Fe, Co, and Ni) heterostructures. Within one group of the periodic table, electronegativity increases with increasing number of *d*-electrons from the left ($LaMnO_3$) to the right ($LaNiO_3$). Energy difference between the local 5*d*- and the 3*d*-states (dashed lines) at the interface increases with *n* leading to enhanced charge transfer. (f) Sketch of the band-line up of a I(metal)/X(insulator)/I(metal) heterostructure. Electron wavefunction Ψ(x) is delocalized in the metallic I-layer and decreases exponentially in the insulating X-layer. In-gap- or surface/interfacial-states may lead to additional electron localization in the X-layer yielding Ψ(x) ≈ const. for *t* = 4 ML.

Mechanism (ii) arises from polarity discontinuities in the [$I_iX_i$] SLs, which can trigger charge transfer from the X- to the I-layer (n-type interface with respect to I) or in the opposite direction (p-type interface) depending on the layer-termination. Compensation of the built-in electric field by electronic reconstruction is expected for layer thicknesses of typically ≳4 MLs and could therefore occur in the SLs.[65] An $IrO_2$-termination of the I-layer yields an n-type interface, whereas a SrO termination results in a p-type interface. For [$I_4X_4$] SLs, the HAADF-STEM analysis reveals SrO-termination of the I-layer (p-type) and LaO-termination of the X-layer, such that the next I-layer starts with $IrO_2$ (n-type), see SI. The layers of the SL therefore contain both types of interfaces, n-type at the bottom and p-type at the top of the SIO layer, which likely compensates built-in electric fields and suppresses ICT driven by mechanism (ii). First-principles studies of 3*d*/5*d* TMO superlattices indicate that polarity becomes important when built-in electric fields are not compensated.[66]

Mechanism (iii) involves differences in electronegativity between the layers (Fig. 4d). The electronegativity or affinity of a TM ion increases with increasing filling of the *d*-shell for $n \geq 5$.[67] In general, the larger the electronegativity difference between two layers, the stronger the charge transfer toward the layer with higher electronegativity. For corner-sharing TMO perovskites, O 2*p*-states align across the interface so that the difference in TM electronegativity translates into a chemical potential discontinuity that drives charge transfer.[25,38,68]

Recently, DFT calculations on intrinsic charge transfer in complex oxide interfaces suggest the energy difference $\Delta\varepsilon_p$ between the O 2*p*-states of neighboring TMOs as the driving energy for the ICT.[38] Because the TM *d*-states near the Fermi level are hybridized with O 2*p*-states, alignment of the O 2*p*-bands across the interface modifies the energy of the local 3*d* states $\varepsilon_d$ and thus the difference $\varepsilon_d - \varepsilon_p$. As a result, electrons transfer toward the side with the higher $\varepsilon_p$, minimizing the overall electronic energy. The local energy of the O 2*p*-states, $\varepsilon_p$, not only increases systematically with increasing filling of the *d*-shell but also with increasing oxidation state of the *B*-site ion.[38] In contrast, within a given group of the periodic table, $\varepsilon_p$ decreases from 3*d* to 5*d* compounds. Based on these considerations, in [$I_4X_4$]$_5$ SLs, a systematic increase of electron transfer is expected from Ir ($5d^5$) to X= Mn ($3d^4$), Fe ($3d^5$), and Co ($3d^6$), see Fig. 4e. Experimentally, this trend is observed in [$I_4X_4$]$_5$ SLs, where the charge transfer increases with 3*d*-shell filling. In line with these results, iridate SLs containing $LaNiO_3$, *i. e.*, Ni ($3d^7$), indeed exhibit large charge transfer to the Ni-site.[37] This strongly supports scenario (iii) as the dominant mechanism for ICT. The metallic character of the I-layer and electron localization in the X-layer provide spatially homogeneous distribution of transferred charge carriers in the confined layers, see Fig. 4f.

More quantitatively, DFT predicts a linear relationship between the transferred charge $\Delta n_e$ and the bulk O 2*p* energy $\varepsilon_p$ of the TMOs.[38] In Fig. 5a we plot $\Delta n_e$ per ion, as deduced from the Ir XAS data versus $\varepsilon_p$-$E_F$. The measured data follow the trend expected from the oxygen-2*p*-level alignment scenario. While the oxygen-2*p* energies are not measured directly here, the observed charge-transfer direction and systematic evolution across the Mn-, Fe-, and Co-based superlattices support electronegativity mismatch as the dominant contribution to ICT in this series. The charge transfer for a $SrIrO_3$/$LaNiO_3$ SL is estimated to be $\Delta n_e \approx 0.5$, obtained by linear approximation to $\varepsilon_p$-$E_F$ = -0.75 eV. The value of $\varepsilon_p$-$E_F$ for $LaNiO_3$ was extracted from Fig. S7 (see SI). However, for $SrIrO_3$/$NdNiO_3$ a reduced charge transfer has been reported,[69] indicating that different mechanisms may be active in iridate-nickelate SLs. In the rare-earth (*R*) nickelates $R$$NiO_3$, the O 2*p*-states lie very close to, or even above, the energy of the 3*d*-states, resulting in strong Ni-O hybridization,[70] which can be tuned by the *R*-site ionic radius, *i. e.*, chemical pressure.[22] With decreasing *R*-size, the Ni-O bond angle, and thus the hybridization, decreases in $RNiO_3$ perovskites. The resulting reduction of Ni-O hybridization and covalency lowers the effective bond electronegativity,[68,71,72] which explains the reduced charge transfer in $SrIrO_3/NdNiO_3$ compared to $SrIrO_3/LaNiO_3$ heterostructures.

Similar to chemical pressure, epitaxial strain may also significantly affect TM 3*d*-O 2*p* hybridization and hence electronegativity. For example, 1% of compressive strain lowers $\varepsilon_p$ in $SrMnO_3$ by about 10%.[38] Hence, epitaxial strain may also contribute to ICT. For the X-layers grown on (001) STO, the M and F layers are under compressive strain, whereas C layers experience tensile strain. An opposite effect on $\varepsilon_p$ would therefore be expected, which, however, is not observed, or at least appears negligible, considering Fig. 5a.

Finally, we address the issue of A-site intermixing of Sr and La at the interface. In case of charge-transfer mechanism (i), ICT is expected to be reduced, whereas for scenario (ii), the influence is likely canceled due to the presence of both n-type and p-type interfaces. For mechanism (iii), intermixing of up to about 25% can still be regarded as a secondary effect with respect to ICT and should not significantly alter charge transfer.[38]

The experimental results indicate that ICT in $[I_iX_i]_m$-type SLs is dominantly driven by differences in electronegativity, effectively diminishing or compensating discontinuities of the chemical potential across the interface. Interfacial band alignment and hybridization evidently modify not only band-filling but also the electronic states. The $e_g$ related $3dz^2$-$r^2$ orbitals, which are aligned along the *c*-axis (perpendicular to the interface) display the largest overlap between 5*d* and 3*d* states and are therefore particularly favored for the formation of molecular orbitals.[73] Hybridization of $e_g$ -orbitals lowers $e_g$ - $t_{2g}$ energy splitting in the X-layer, which in the case of X = C (see Fig. 5b,c) provides a simple explanation for a LS to HS conversion. For X = M and F, such effects are not visible in the BR owing to the robust HS state.[54–57]

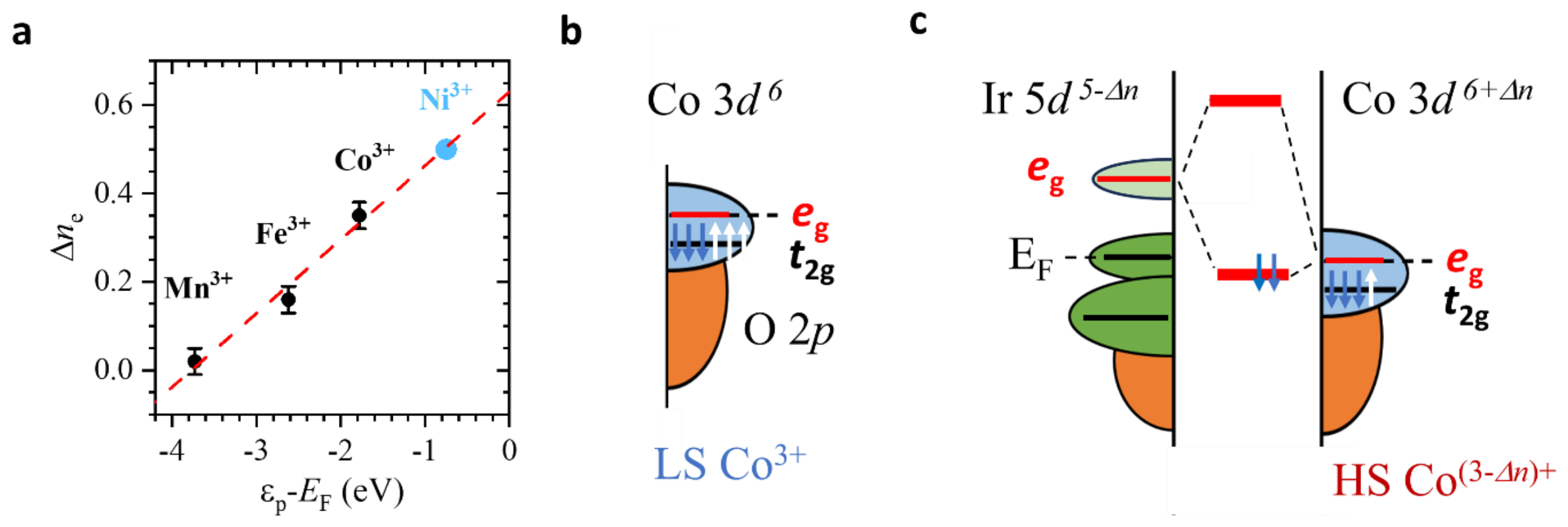

**Figure 5**. (a) The electron transfer $\Delta n_e$ per ion from the I-layer versus $\varepsilon_p$-$E_F$ of bulk M, F, and C (black symbols) with tri-valent TMs. $\Delta n_e$ for $LaNiO_3$ (blue symbol) was obtained by a linear approximation of $\varepsilon_p$-$E_F$ vs. electron filling *n* of the 3*d* shell (see SI). Values for $\varepsilon_p$-$E_F$ are taken from Ref.[38] Schematic of the Co-O bands close to $E_F$ before (b) and after hybridization (c) at the I-C interface. Energy lowering of $e_g$ states stabilizing HS configuration of the Co ion ($2e_g$ $4t2_g$) after hybridization which otherwise is dominantly in LS configuration ($6t2_g$).

## 3. Conclusion

Charge transfer in $5d/3d$ TMO SLs was studied in detail by XAS and EELS analysis. Charge transfer always occurs from the $5d$ I-layer to the $3d$ X-layer, consistent with the larger electronegativity of the $3d$ TMOs. Electron depletion in I and accumulation in X are equal in magnitude and homogeneously distributed within the short-period $[I_4X_4]_5$ SLs, enabling a quantitative comparison of ICT for different X. Among the most prominent mechanisms proposed for ICT in TMO heterostructures, charge transfer is best described by differences in electronegativity. The transferred charge across the interface per ion, $\Delta n_e$, quantitatively matches the expectation based on the electronegativity difference between the layers. A maximum of $\Delta n_e \approx 0.35e$ is observed for the $[I_4C_4]_5$ SL. Beyond interfacial doping, band alignment can also lead to significant band hybridization, thereby modifying the electronic configuration. The enhanced branching ratio for X=C indicates distinct LS to HS conversion, which is likely related not only to epitaxial strain but also to the formation of hybridized states at the interface.

Overall, the presented experimental results establish electronegativity mismatch as a predictive design parameter for TMO heterostructures. This approach enables not only tuning of band-filling but also control over the energy of electronic states, providing prospects for advanced oxide electronics and next-generation information technologies.

## 4. Experimental Section

*Sample Preparation:* The preparation of epitaxial superlattices (SLs) was carried out by pulsed laser deposition (PLD) from stoichiometric targets on (001) oriented $TiO_2$-terminated STO substrates.[74] The film growth was monitored by in-situ high pressure reflection high energy electron diffraction (RHEED), documenting a layer-by-layer growth mode for all the films and allowing thickness control of the deposition on the scale of one ML.[18] For SLs of the form $[I_iX_i]_m$ we first deposited an I-layer with a thickness of *i*-MLs followed by a X-layer with same thickness. The modulation was repeated *m*-times. After film deposition, the samples were post-annealed at 500°C in 100 mbar $O_2$ for about 0.5 h to provide full oxygenation.

*Structural Characterization:* X-ray analysis was carried out at ambient conditions with a lab source Bruker D8 Davinci diffractometer equipped with Cu $K_\alpha$ radiation ($\lambda$ = 1.5418 Å). The cross-sectional TEM specimens of the $[I_4X_4]_5$ SLs were prepared by focused ion beam (FIB) (Strata dual beam, FEI Company). The interface structure at atomic resolution were imaged by a Thermo Fisher Scientific Themis Z electron microscope, operating at 300kV and equipped with both image and probe correctors. The atomically resolved elemental maps were acquired by STEM-EDX spectrum imaging using a Super-X detector. The atomically resolved EELS spectrum imaging (SI) was acquired on the same microscope in STEM mode with Gatan Continuum K3 1069HR image filter. SI were acquired in dual EELS mode, i.e., for each pixel a zero loss and a high loss EEL spectra were acquired to calibrate the energy level of the core loss edges. Zero loss was also used to remove the plural scattering from the high loss edge by Fourier-ratio processing. The energy losses corresponding to the respective transition metals were applied by Drift Tube with 520eV for Mn, 690eV for Fe and 750 eV for Co. For all the specimens, the step size is 0.09 nm, convergence angle being 30 mrad and collection angle 40 mrad. The energy dispersion was set to 0.09 eV/channel. *X-ray Absorption Spectroscopy:* The X-ray absorption spectra (XAS) at the Ir $L_{2,3}$-edges were measured at the beamline ID12 at the European Synchrotron Radiation Facility (ESRF) in Grenoble, France. All the spectra were recorded with left- and right circular polarized photons using the second harmonic of the HU-38 undulator. The samples were mounted on a sample holder making an angle of 15° with respect to the incident X-rays. Measurements were performed at $T$ = 20 K using partial fluorescence yield detection mode with two energy resolved detectors mounted in backscattering geometry.

*Electronic Transport:* The electronic transport of the samples was characterized by four-point resistance measurements in Van der Pauw geometry using a standard physical property measurement system (PPMS) equipped with a 14 T superconducting solenoid. Contacts to the SLs were prepared by ultrasonic Al-wire bonding to the corners of the 5×5 $mm^2$ film surface. Resistance measurements were done by using alternating current with frequency and amplitude of 2Hz and 0.1 mA, respectively.

## Supporting Information

Supporting Information is available from the Wiley Online Library or from the author.

## Acknowledgements

A.K.J. acknowledges financial support from the European Union's Framework Programme for Research and Innovation, Horizon 2020, under the Marie Skłodowska-Curie grant agreement No. 847471 (QUSTEC). D.W. acknowledges the support from collaborative research centre FLAIR (Fermi level engineering applied to oxide electro-ceramics), which is funded by the German Research Foundation (DFG), project-ID 463184206 – SFB 1548. D.F. and A.K.J. thank Robert Eder and Amir Haghighirad from the IQMT and Tanusri Saha-Dasgupta and Samir Rom from the S. N. Bose National Centre for Basic Science (Kolkata, India) for fruitful discussions.

## Conflict of Interest

The authors declare no conflict of interest.

## Data Availability Statement

The data that support the findings of this study are available from the corresponding author upon reasonable request.

## Author Contributions

A.K.J. and D.W. contributed equally to this work. D.F. conceived and designed the research. A.K.J. designed sample preparation, structural characterization and transport measurements. J.L. and S.R. equally supported sample preparation and experiments. D.W. and V.W. carried out TEM sample preparation and analysis. A.K.J., D.F., F.W., and A.R. performed XAS measurements. D.F. and M.L.T. supervised the project and wrote the manuscript with input from all coauthors.